\documentclass[a4paper,twoside,12pt]{article}
\input{obsjkt.sty}

\newcommand{\Msun}{\ensuremath{\,{\rm M}_\odot}}                  % Solar mass symbol
\newcommand{\Rsun}{\ensuremath{\,{\rm R}_\odot}}                  % Solar radius symbol
\newcommand{\rhosun}{\ensuremath{\,\rho_\odot}}                   % Solar density symbol
\newcommand{\Teff}{\ensuremath{T_{\rm eff}}}                      % Effective temperature symbol
\newcommand{\Vsys}{\ensuremath{V_\gamma}}                         % Systemic velocity symbol
\newcommand{\degr}{\ensuremath{^\circ}}                           % Degree symbol
\renewcommand{\kms}{\,km\,s$^{-1}$}                               % km/s symbol
\newcommand{\etal}{\textit{et al.}}                               % et al. in italics

\newcommand{\tess}{\textit{TESS}}
\newcommand{\hip}{\textit{Hipparcos}}
\newcommand{\gaia}{\textit{Gaia}}

\newcommand{\Msunnom}{\hbox{$\mathcal{M}^{\rm N}_\odot$}}
\newcommand{\Rsunnom}{\hbox{$\mathcal{R}^{\rm N}_\odot$}}
\newcommand{\Lsunnom}{\hbox{$\mathcal{L}^{\rm N}_\odot$}}

\begin{document} %%%%%%%%%%%%%%%%%%%%%%%%%%%%%%%%%%%%%%%%%%%%%%%%%%%%%%%%%%%%%%%%%%%%%%%%%%%%%%%%%%%%%%%%%%%%%%%%%%%%%%%%%%%%%%%%%%%%%%%%%%%%%%%%%%%%
%%%%%%%%%%%%%%%%%%%%%%%%%%%%%%%%%%%%%%%%%%%%%%%%%%%%%%%%%%%%%%%%%%%%%%%%%%%%%%%%%%%%%%%%%%%%%%%%%%%%%%%%%%%%%%%%%%%%%%%%%%%%%%%%%%%%%%%%%%%%%%%%%%%%%

\OBSheader{Rediscussion of eclipsing binaries: V505 Per}{J.\ Southworth}{2021 Oct}

\OBStitle{Rediscussion of eclipsing binaries. Paper VI. \\ The F-type system V505 Persei}

\OBSauth{John Southworth}

\OBSinstone{Astrophysics Group, Keele University, Staffordshire, ST5 5BG, UK}

\OBSabstract{V505\,Per is a detached eclipsing binary containing two F5\,V stars in a 4.22\,d circular orbit. We use a light curve from the \tess\ satellite and published radial velocity measurements to establish the properties of the system to high precision. The masses of the stars are $1.275 \pm 0.004$\Msun\ and $1.258 \pm 0.003$\Msun, and their radii are $1.294 \pm 0.002$\Rsun\ and $1.264 \pm 0.002$\Rsun. Adding published effective temperature estimates, we precisely measure the luminosities and absolute bolometric magnitudes of the stars, and the distance to the system. The distance is slightly shorter than that obtained from the \gaia\ EDR3 parallax, a discrepancy most easily explained by uncertainty in the 2MASS $K$-band apparent magnitude. We reanalyse existing light and radial velocity curves from three previous studies of this system and conclude that, in this case, formal errors are reliable for the spectroscopic orbits but not light curves, that errorbars from a residual-permutation algorithm are suitable for light curves but not spectroscopic orbits, and that published results are not always reproducible. The precisions in the measured properties of V505\,Per are high and among the best ever obtained for a detached eclipsing binary system.}

%%%%%%%%%%%%%%%%%%%%%%%%%%%%%%%%%%%%%%%%%%%%%%%%%%%%%%%%%%%%%%%%%%%%%%%%%%%%%%%%%%%%%%%%%%%%%%%%%%%%%%%%%%%%%%%%%%%%%%%%%%%%%%%%%%%%%%%%%%%%%%%%%%%%%

\section*{Introduction}

Detached eclipsing binaries (dEBs) offer the possibility of measuring the physical properties of stars to high precision and accuracy without any reliance on theoretical models of stellar evolution, so represent a direct probe of how stars evolve \cite{Russell14nat} and an important testbed for the verification and refinement of theoretical models \cite{Andersen++90apj,Pols+97mn,Spada+13apj,ClaretTorres18apj,Tkachenko+20aa}. High precision and accuracy in the measured mass and radius is vital for this work, as is a precise measurement of the effective temperature (\Teff) and chemical composition of each star \cite{Torres++10aarv,Valle+18aa}. The reliability of mass and radius determinations can be assessed by comparison of the results from multiple independent analyses of individual or different datasets for the same dEB \cite{Me+05mn,Maxted+20mn}. 

In this work we present a detailed analysis of the dEB V505\,Persei, which consists of two very similar F5\,V stars. Our analysis is based on three published radial velocity (RV) studies, three published light curves and a new light curve obtained by the NASA Transiting Exoplanet Survey Satellite \cite{Ricker+15jatis} (\tess) mission. This forms part of our project to systematically revise and improve the measured physical properties of known dEBs \cite{Me20obs,Me21obs1,Me21obs2,Me21obs3,Me21obs4}, in particular those which can be included in DEBCat\footnote{\texttt{https://www.astro.keele.ac.uk/jkt/debcat/}}, a catalogue of dEBs with mass and radius measurements to precisions of 2\% or better \cite{Me15debcat}.

%%%%%%%%%%%%%%%%%%%%%%%%%%%%%%%%%%%%%%%%%%%%%%%%%%%%%%%%%%%%%%%%%%%%%%%%%%%%%%%%%%%%%%%%%%%%%%%%%%%%%%%%%%%%%%%%%%%%%%%%%%%%%%%%%%%%%%%%%%%%%%%%%%%%%

\begin{table}[t]
\caption{\em Basic information on V505\,Per \label{tab:info}}
\centering
\begin{tabular}{lll}
{\em Property}                 & {\em Value}            & {\em Reference}                \\[3pt]
% Bright Star Catalogue        & HR NNNN                & \cite{HoffleitJaschek91}       \\
Henry Draper designation       & HD 14384               & \cite{CannonPickering18anhar}  \\
\textit{Hipparcos} designation & HIP 10961              & \cite{Hipparcos97}             \\
\textit{Tycho} designation     & TYC 3690-536-1         & \cite{Hog+00aa}                \\
\textit{Gaia} EDR3 designation & 455772347387763840     & \cite{Gaia20aa}                \\
\textit{Gaia} EDR3 parallax    & $16.069 \pm 0.020$ mas & \cite{Gaia20aa}                \\
\tess\ designation             & TIC 348517784          & \cite{Stassun+19aj}            \\
$B$ magnitude                  & $7.30  \pm 0.01 $      & \cite{Hog+00aa}                \\
$V$ magnitude                  & $6.88  \pm 0.01 $      & \cite{Hog+00aa}                \\
$J$ magnitude                  & $6.070 \pm 0.067$      & \cite{Cutri+03book}            \\
$H$ magnitude                  & $5.793 \pm 0.036$      & \cite{Cutri+03book}            \\
$K_s$ magnitude                & $5.771 \pm 0.020$      & \cite{Cutri+03book}            \\
Spectral type                  & F5\,V + F5\,V          & \cite{Tomasella+08aa}          \\[10pt]
\end{tabular}
\end{table}

\section*{V505\,Persei}

In this work we present an analysis of the dEB V505\,Per (Table\,\ref{tab:info}) based on its light curve from \tess\ and on published RVs. V505\,Per is an F-type system containing two very similar stars on a circular orbit with a period of 4.22\,d. The discovery was announced by Kaiser \cite{Kaiser89jaavso} under the guise of SAO\,23229 and with a period of 2.111\,d, half the true value. Kaiser \etal\ \cite{Kaiser++90ibvs} presented nine times of minimum light estimated visually and established the first ephemeris; they noted that their period of 2.1110084\,d %(12)\,d\footnote{In this work errorbars are sometimes given in brackets. In this case the number in brackets refers to an errorbar stripped of any leading zeros.} 
might be half the true value if the primary and secondary minima were of similar depth (as indeed turned out to be the case).

Marschall \etal\ \cite{Marschall+90ibvs} obtained spectroscopy and found that the system was double-lined and with an orbital period of 4.22\,d. Marschall \etal\ \cite{Marschall+97aj} (hereafter MA97) presented a detailed study of the system based on 63 \'echelle spectra and a light curve comprising 1324 datapoints in the $B$ and $V$ filters. They determined the masses of the stars to high precision, but their radius measurements were good to only 2.3\% (primary, hereafter star~A) and 5.5\% (secondary, hereafter star~B) due to the scatter in their photometry as well as the intrinsic indeterminacy of the ratio of the radii of a dEB showing deep but partial eclipses. 

Munari \etal\ \cite{Munari+01aa} (hereafter MU01) studied V505\,Per in the context of investigating what might be achieved using \gaia\ photometry and spectroscopy for dEBs. RVs were obtained from 20 \'echelle spectra of the calcium infrared triplet (849.8, 854.2 and 866.2 nm) and the \hip\ light curve of this object. The mass measurements were less precise than those of MA97, as they were deliberately based on data of lower quality in order to mimic \gaia. The radii were measured to much higher precision (1.4\% for star~A and 2.6\% for star~B) despite being obtained from a light curve with only 11 datapoints during eclipse. The authors did note that this is a \emph{formal} error but made no attempt to determine a true uncertainty.

Tomasella \etal\ \cite{Tomasella+08aa} (hereafter TO08) reanalysed V505\,Per based on 36 new \'echelle spectra and 627 light curve datapoints in $B$ and $V$. They obtained mass measurements in good agreement with those of MA97, but radius measurements with much smaller errorbars (less than 1\%) despite the similarity of the light curves presented by the two works. They also measured the atmospheric parameters of the stars (\Teff\ and [M/H]) via a $\chi^2$ analysis of the spectra -- the use of the $\chi^2$ statistic on observations with significant modelling uncertainties (e.g.\ high-resolution theoretical spectra \cite{Kurucz03conf,Smalley05msais}) is questionable.

Finally, Baugh \etal\ \cite{Baugh+13pasp} used high-signal \'echelle spectra to measure the photospheric lithium abundance of the components of V505\,Per, both of which are in the lithium dip ($\Teff \sim6400$ to 6800\,K; Refs.\ \cite{Jeffries14conf,Lyubimkov16ap}). They found that its lithium was less depleted than expected for its age, confirming the hypothesis that the different rotational evolution of stars in short period binaries affects their lithium depletion.

%%%%%%%%%%%%%%%%%%%%%%%%%%%%%%%%%%%%%%%%%%%%%%%%%%%%%%%%%%%%%%%%%%%%%%%%%%%%%%%%%%%%%%%%%%%%%%%%%%%%%%%%%%%%%%%%%%%%%%%%%%%%%%%%%%%%%%%%%%%%%%%%%%%%%

\section*{Observational material}

\begin{figure}[t] \centering \includegraphics[width=\textwidth]{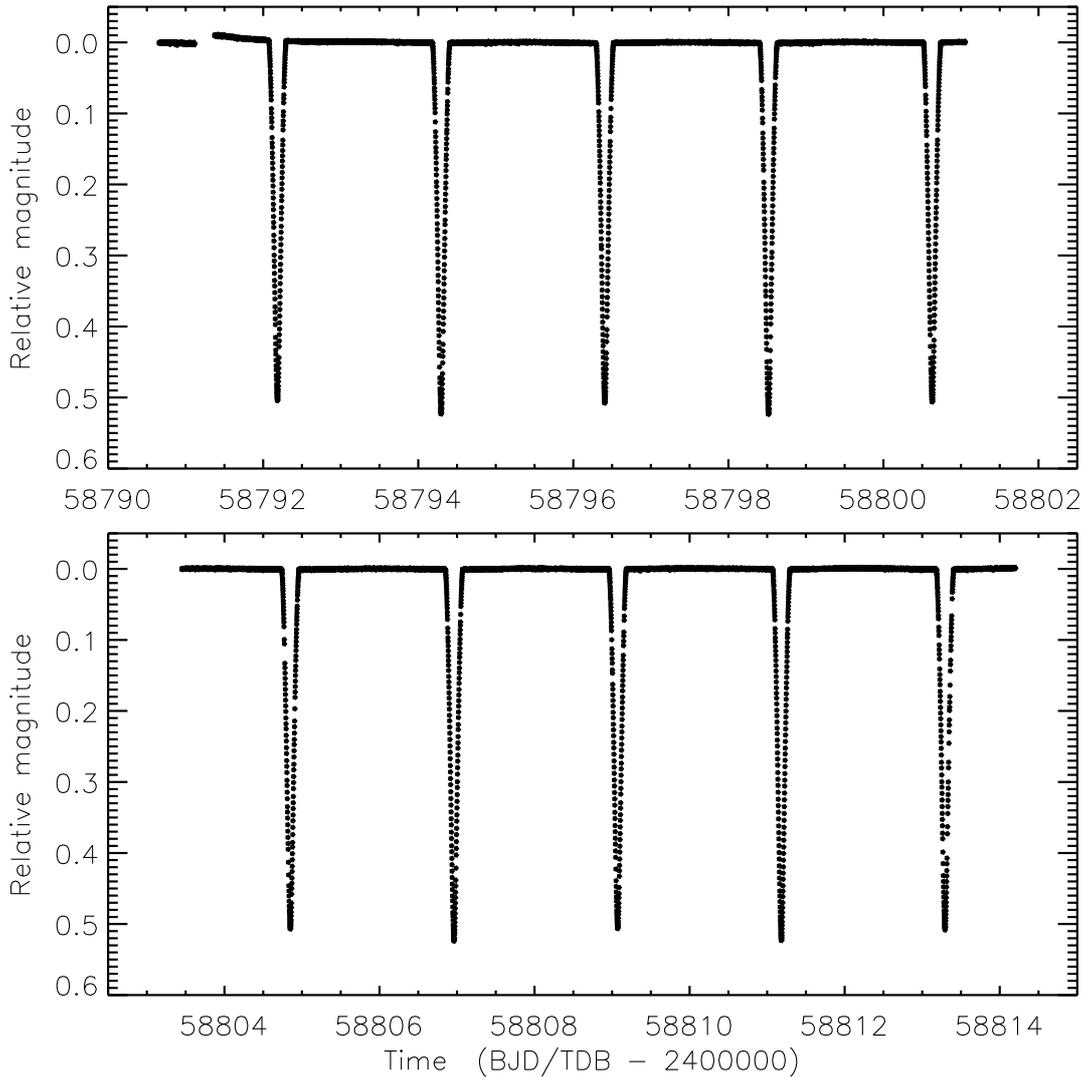} \\
\caption{\label{fig:time} \tess\ Sector 18 short-cadence photometry of V505\,Per.
The two panels show the data before and after the mid-sector pause for download
of the data to Earth \cite{Ricker+15jatis}.} \end{figure}

% Sector 18 (2019-Nov-02 to 2019-Nov-27, in cycle 2): observed in camera 2.

In this work we concentrate on the light curve of V505\,Per obtained using the NASA \tess\ satellite \cite{Ricker+15jatis}, which observed it in Sector 18 (2019/11/02 to 2019/11/27). The observations cover four primary and six secondary eclipses, with one primary eclipse lost to the mid-sector pause for downlinking the data from the satellite to Earth (Fig.\,\ref{fig:time}). These data were downloaded from the MAST archive\footnote{Mikulski Archive for Space Telescopes, \\ \texttt{https://mast.stsci.edu/portal/Mashup/Clients/Mast/Portal.html}} and converted to relative magnitude. We retained only those datapoints with the QUALITY flag equal to zero. 

The \tess\ data were obtained in short cadence mode, with a sampling rate of 120\,s. We chose to use the simple aperture photometry (SAP) data rather than the pre-search data conditioning (PDC) alternative \cite{Jenkins+16spie}. This is because the PDC data are processed with the aim of finding shallow planetary transits, an approach that often introduces artefacts in light curves of stars such as dEBs with a strong intrinsic variability. Of the 17\,554 datapoints, 14\,805 have a QUALITY flg of zero and were retained for further analysis.

%%%%%%%%%%%%%%%%%%%%%%%%%%%%%%%%%%%%%%%%%%%%%%%%%%%%%%%%%%%%%%%%%%%%%%%%%%%%%%%%%%%%%%%%%%%%%%%%%%%%%%%%%%%%%%%%%%%%%%%%%%%%%%%%%%%%%%%%%%%%%%%%%%%%%

\section*{Analysis of the \tess\ light curve}

\begin{figure}[t] \centering \includegraphics[width=\textwidth]{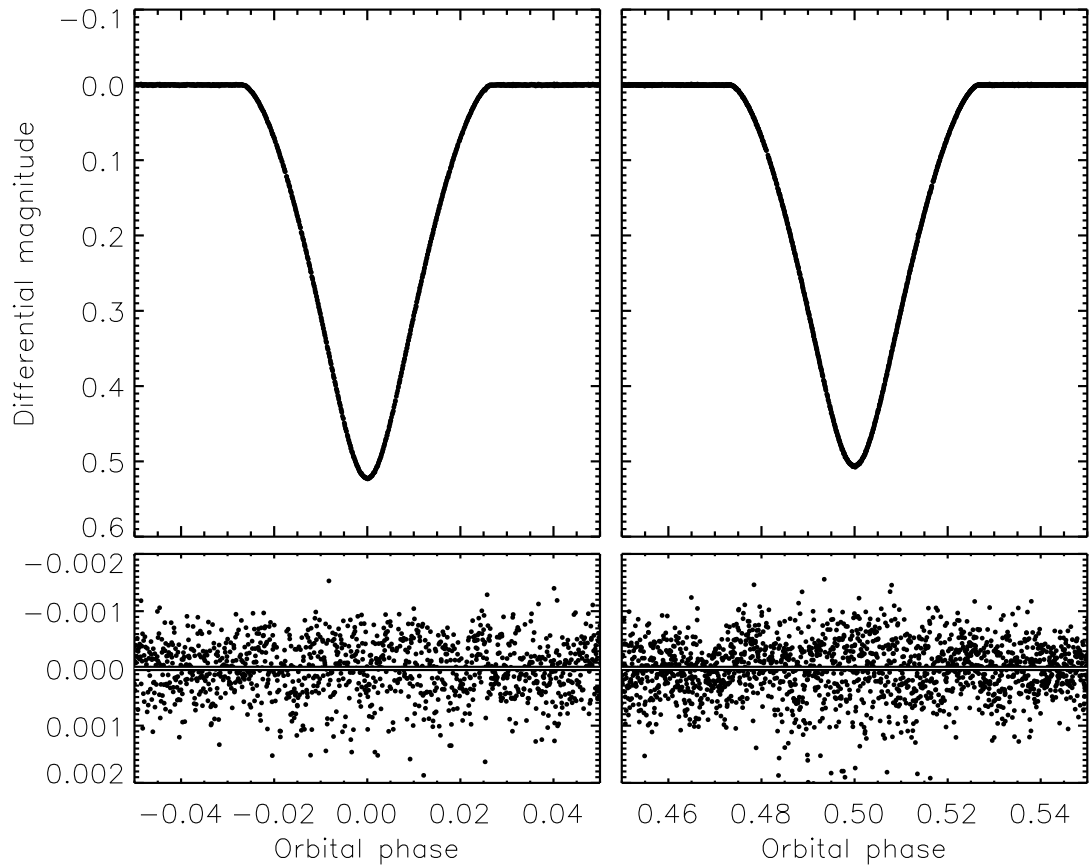} \\
\caption{\label{fig:tess} The \tess\ light curve of V505\,Per (filled circles) around the primary (left) and 
secondary (right) eclipses. The best fit is not plotted as it is indistinguishable from the data. The lower 
panels show the residuals of the fit with the line of zero residual overplotted in white for clarity.} \end{figure}

We first analysed the \tess\ photometry of V505\,Per in order to establish the most reliable photometric parameters of the system. For this we used version 41 of the {\sc jktebop}\footnote{\texttt{http://www.astro.keele.ac.uk/jkt/codes/jktebop.html}} code \cite{Me++04mn2,Me13aa}, which is suitable for systems with well-separated stars \cite{Maxted+20mn}. We used the definition that the primary eclipse is deeper than the secondary eclipse, and set star~A to be the star eclipsed during primary eclipse. The two stars are very similar, but star~A is slightly hotter, larger and more massive than star~B.

To save computation time and to avoid problems with slow changes in the brightness of the system due to instrumental effects, we extracted from the \tess\ light curve every datapoint within one eclipse duration of the midpoint of an eclipse. We fitted for the sum ($r_{\rm A}+r_{\rm B}$) and ratio ($k = \frac{r_{\rm B}}{r_{\rm A}}$) of the fractional radii, defined by $r_{\rm A} = \frac{R_{\rm A}}{a}$ and $r_{\rm B} = \frac{R_{\rm B}}{a}$ where $R_{\rm A}$ and $R_{\rm B}$ are the true radii of the stars, and $a$ is the semimajor axis of the relative orbit. We also fitted for the orbital inclination ($i$) and ephemeris ($P$ and $T_0$), and for the central surface brightness ratio of the two stars ($J$). Limb darkening was included using the quadratic law \cite{Kopal50} with theoretical coefficients from Claret \cite{Claret18aa}. The same limb darkening coefficients were used for both stars, a reasonable step due to their similarity: the linear coefficient was fitted and the quadratic coefficient was fixed. Third light was held at zero because fits with it included as a free parameter returned almost identical results with an insignificant and negative amount of third light. The orbit was also assumed to be circular as we found no evidence for eccentricity. The other fitted parameters were the coefficients of nine straight lines applied to the out-of-eclipse brightness of the system, one for each eclipse.

\begin{table} \centering
\caption{\em \label{tab:lc} Parameters of the best {\sc jktebop} fit to the \tess\ light curve of V505\,Per.
The uncertainties are 1$\sigma$. The same limb darkening coefficients were used for both stars.}
\begin{tabular}{lr@{\,$\pm$\,}l}
{\em Parameter}                           & \multicolumn{2}{c}{\em Value}    \\[3pt]
{\it Fitted parameters:} \\
Primary eclipse time (BJD/TDB)            & 2458798.516720  & 0.000005        \\
Orbital period (d)                        &      4.2220216  & 0.0000023       \\
Orbital inclination (\degr)               &      87.9166    & 0.0030          \\
Sum of the fractional radii               &       0.170906  & 0.000035        \\
Ratio of the radii                        &       0.9788    & 0.0019          \\
Central surface brightness ratio          &       0.97775   & 0.00050         \\
Third light                               & \multicolumn{2}{c}{~~~~~~~~~~0.0 (fixed)}   \\
Linear limb darkening coefficient         &       0.261     & 0.005            \\
Quadratic limb darkening coefficient      & \multicolumn{2}{c}{~~~~~~~~~~0.23 (fixed)}  \\[5pt]
% $e\cos\omega$                             &    $-$0.006803  & 0.000007        \\
% $e\sin\omega$                             &       0.00712   & 0.00037         \\
{\it Derived parameters:} \\
Fractional radius of star~A               &       0.086370  & 0.000078        \\
Fractional radius of star~B               &       0.084536  & 0.000091        \\
% Orbital eccentricity                      &       0.00985   & 0.00027         \\
% Argument of periastron (\degr)            &     133.7       & 1.5             \\
Light ratio                               &       0.9367    & 0.0037          \\
\end{tabular}
\end{table}
    
The best fit is shown in Fig.\,\ref{fig:tess} and is a very good description of the data; the scatter around the best fit is only 0.49\,mmag. The parameter values determined are given in Table\,\ref{tab:lc}. The residuals have a slightly non-Gaussian distribution, with a longer tail to fainter magnitudes. This effect is typical in \tess\ light curves (e.g.\ Refs.\ \cite{Me21obs2} and \cite{Me21obs4}). 

The uncertainties in the fitted parameters were determined in three ways: Monte Carlo (MC) and residual-permutation (RP) algorithms \cite{Me++04mn,Me08mn} and by splitting the data into three subsets each containing three consecutive eclipses. For each parameter we adopted the larger of the MC and RP alternatives; we used the third method only as a consistency check due to the small-number statistics intrinsic to the current case. In Paper\,V (Ref. \cite{Me21obs4}) we found that the three error estimates agreed well, and thus were likely to be reliable even in cases where there were significant correlations between some parameters. We found that the RP uncertainties were larger than the MC uncertainties by typically 50\%, so these were adopted as our final uncertainties. 

% v505per_sap_ecl_01.par:Fractional primary radius:                 0.0863151172
% v505per_sap_ecl_02.par:Fractional primary radius:                 0.0861924798
% v505per_sap_ecl_03.par:Fractional primary radius:                 0.0866950473
% v505per_sap_ecl_01.par:Fractional secondary radius:               0.0846009392
% v505per_sap_ecl_02.par:Fractional secondary radius:               0.0847367815
% v505per_sap_ecl_03.par:Fractional secondary radius:               0.0841292228
% r1=[0.0863151172,0.0861924798,0.0866950473]
% r2=[0.0846009392,0.0847367815,0.0841292228]
% print, mean(r1), stddev(r1)/sqrt(3) &$
% print, mean(r2), stddev(r2)/sqrt(3)

%%%%%%%%%%%%%%%%%%%%%%%%%%%%%%%%%%%%%%%%%%%%%%%%%%%%%%%%%%%%%%%%%%%%%%%%%%%%%%%%%%%%%%%%%%%%%%%%%%%%%%%%%%%%%%%%%%%%%%%%%%%%%%%%%%%%%%%%%%%%%%%%%%%%%

\section*{Analysis of published light curves}

Three previous works have studied V505\,Per using a variety of light curves and with a range of approaches to understanding the uncertainties in the derived parameters. We therefore attempted to reproduce these results. We were unfortunately not able to obtain the light curves from MA97 so these were discluded from our analysis.

MU01 used the \hip\ and \textit{Tycho} light curves \cite{Hipparcos97,Hog+00aa} to evaluate the results potentially achievable for dEBs using the \gaia\ satellite. We obtained these data\footnote{\texttt{http://vizier.u-strasbg.fr/viz-bin/VizieR?-source=I/239/hip\_main\&recno=10953}} and analysed the \hip\ observations. We did not include the \textit{Tycho} data as these are much more scattered than the \hip\ magnitude measurements so contribute negligible additional information. The data were modelled using {\sc jktebop} with $r_{\rm A}$, $r_{\rm B}$, $i$, $J$, $T_0$ and out-of-eclipse brightness as the fitted parameters. We fixed the limb darkening coefficients to theoretical values \cite{ClaretHauschildt03aa} for the $V$-band.  The orbital period was fixed to the value given in Table\,\ref{tab:lc}. Uncertainties were obtained using MC and RP simulations. The results are compared in Table\,\ref{tab:lcmany} to those from MU01 for $r_{\rm A}$, $r_{\rm B}$ and $i$. We find consistent values but with uncertainties larger by factors of between 2 and 4. Uncertainties in $r_{\rm A}$ and $r_{\rm B}$ were not given by MU01 so we have used the (fractional) uncertainties for the true radii of the stars and neglected the much smaller contribution to these from the uncertainty in the semimajor axis. We confirm that the uncertainties found by MU01 are underestimated.

%             r1                      r2                      i
%                   
% Marschall   0.0861      0.0022      0.0844      0.0043      87.83 0.02
% 
% Munari:     0.09297     0.00133     0.07570     0.00199     88.18 0.11
% Me (Hip):   0.09143     0.00493     0.07823     0.00585     88.09 0.22
% 
% Tomasella   0.0860      0.0009      0.0846      0.0009      87.95 0.04
% Me (Tom B)  0.0901      0.0017      0.0833      0.0021      87.90 0.14        MC/MC/RP
% Me (Tom V)  0.0914      0.0029      0.0816      0.0046      87.89 0.27        RP/RP/RP
% 
% Me (TESS)   0.08637     0.00008     0.08454     0.00009     87.917 0.003

\begin{table} \centering
\caption{\em \label{tab:lcmany} Comparison of measured fractional radii and 
orbital inclination for different analyses and different datasets for V505\,Per.}
\begin{tabular}{@{}lccc}     
{\em Source}              &      $r_{\rm A}$      &      $r_{\rm B}$      &   $i$ ($^\circ$)   \\[5pt]
MA97                      &  $0.0861 \pm 0.0022$  &  $0.0844 \pm 0.0043$  &  $87.83 \pm 0.02$  \\[7pt]
MU01                      &  $0.0930 \pm 0.0013$  &  $0.0757 \pm 0.0020$  &  $88.18 \pm 0.11$  \\
This work (\hip\ data)    &  $0.0914 \pm 0.0049$  &  $0.0782 \pm 0.0059$  &  $88.09 \pm 0.22$  \\[7pt]
TO08                      &  $0.0860 \pm 0.0009$  &  $0.0846 \pm 0.0009$  &  $87.95 \pm 0.04$  \\   
This work (TO08 $B$ data) &  $0.0901 \pm 0.0017$  &  $0.0833 \pm 0.0021$  &  $87.90 \pm 0.14$  \\            % MC/MC/RP
This work (TO08 $V$ data) &  $0.0914 \pm 0.0029$  &  $0.0816 \pm 0.0046$  &  $87.89 \pm 0.27$  \\[7pt]       % RP/RP/RP
This work (TESS data)     & $0.08637 \pm 0.00008$ & $0.08454 \pm 0.00009$ & $87.917 \pm 0.003$ \\[5pt]
\end{tabular}
\end{table}

TO08 analysed their own $BV$ photometry, to which they added the $BV$ photometry from MA97 obtained during eclipse, using the Wilson-Devinney code \cite{WilsonDevinney71apj,Wilson79apj}. Our own analysis of these data necessarily omits the MA97 photometry so is not directly comparable. We modelled the TO08 $BV$ light curves separately using the same approach as in the previous paragraph. We find significantly different results (Table\,\ref{tab:lcmany}): those from TO08 agree well with the (presumed) definitive values from TESS whereas our own analysis of the TO08 data do not. After extensive investigation can only explain this as due to our inability to include the MA97 data. Our uncertainties are significantly larger, and we attribute this to differences in the datasets plus the apparent reliance by TO08 on formal errors computed by the Wilson-Devinney code. Formal errors are known to underestimate the true uncertainties of the fitted parameters \cite{Me20obs,MaceroniRucinski97pasp,Pavlovski+09mn} and should not be used \cite{WilsonVanhamme04}.

%%%%%%%%%%%%%%%%%%%%%%%%%%%%%%%%%%%%%%%%%%%%%%%%%%%%%%%%%%%%%%%%%%%%%%%%%%%%%%%%%%%%%%%%%%%%%%%%%%%%%%%%%%%%%%%%%%%%%%%%%%%%%%%%%%%%%%%%%%%%%%%%%%%%%

\section*{Analysis of published radial velocities}

\begin{table} \centering
\caption{\em \label{tab:orbits} Spectroscopic orbits obtained from each of the three RV datasets. All quantities are in 
\kms. The three systemic velocities are for the stars combined, star~A and star~B, respectively. The bracketed quantities 
were not given by MU01 but were calculated by the current author from quantities that were.}
\begin{tabular}{lrrrrr}
{\em Source}                & $K_{\rm A}$~ & $K_{\rm B}$~ & ${\Vsys}$~ & ${\Vsys}_{\rm ,A}$~ & ${\Vsys}_{\rm ,B}$~ \\[5pt]
                            
MA97                        &       88.93  &       90.30  &      0.040 \\
                            &   $\pm$0.14  &   $\pm$0.14  & $\pm$0.074 \\
This work (RVs from MA97)   &       88.91  &       90.28  &            &           0.00      &            0.10     \\
Formal errors               &   $\pm$0.14  &   $\pm$0.14  &            &      $\pm$0.08      &       $\pm$0.11     \\
MC errors                   &   $\pm$0.14  &   $\pm$0.14  &            &      $\pm$0.10      &       $\pm$0.11     \\
RP errors                   &   $\pm$0.15  &   $\pm$0.14  &            &      $\pm$0.01      &       $\pm$0.01     \\[7pt]
MU01                        & (89.58)$\!\!$& (90.98)$\!\!$&    $-$0.41 \\
                            &              &              &  $\pm$0.39 \\
This work (RVs from MU01)   &       90.23  &       91.85  &            &        $-$0.65      &         $-$0.52     \\
Formal errors               &   $\pm$0.51  &   $\pm$1.36  &            &      $\pm$0.31      &       $\pm$0.93     \\
MC errors                   &   $\pm$0.52  &   $\pm$1.36  &            &      $\pm$0.31      &       $\pm$0.96     \\
RP errors                   &   $\pm$1.59  &   $\pm$1.82  &            &      $\pm$0.81      &       $\pm$1.78     \\[7pt]
TO08                        &       89.01  &       90.28  &       0.21 \\
                            &   $\pm$0.08  &   $\pm$0.09  &  $\pm$0.02 \\
This work (RVs from TO08)   &       89.27  &       90.35  &            &          0.41       &            0.01     \\
Formal errors               &   $\pm$0.12  &   $\pm$0.16  &            &     $\pm$0.08       &       $\pm$0.11     \\
MC errors                   &   $\pm$0.12  &   $\pm$0.15  &            &     $\pm$0.08       &       $\pm$0.11     \\
RP errors                   &   $\pm$0.25  &   $\pm$0.55  &            &     $\pm$0.21       &       $\pm$0.34     \\[7pt]
Final values                &       89.12  &       90.31  \\
                            &   $\pm$0.09  &   $\pm$0.12  \\ 
\end{tabular}
\end{table}

The three previous detailed studies of V505\,Per (MA97, MU01, TO08) have each presented new RV measurements of the system. It is an obvious goal to combine these and thus obtain the greatest precision in the resulting spectroscopic orbit. We have performed independent fits of each of the datasets for two reasons. First, we wish to combine the different orbits using the velocity amplitudes and one of the three previous studies did not present their own values of these quantities. Second, this presents the opportunity to investigate the reliability of the errorbars obtained using various methods.

To do so we obtained the RVs from the PDF files of the three papers\footnote{The PDF version of MA97 was obtained from the NASA ADS website and appears to be an image of the original paper. Selectable text is embedded within the file but seems to have been assembled via optical character recognition software. On cross-checking the datafile with the original paper it was found that a lot of the `5's had been misidentified as `3's.} and fitted them each with {\sc jktebop}. We fixed the orbital period to the value in Table\,\ref{tab:lc} and fitted for the velocity amplitude and systemic velocity of each star plus a phase offset. We assumed a circular orbit and scaled the errorbars of each dataset in order to force a reduced $\chi^2$ of $\chi^2_\nu = 1$. Errorbars were obtained using the MC and RP approaches. Whilst the MC algorithm should be suitable for this work, the RP algorithm may not be. This is because the precision of RVs depends on \emph{orbital phase} rather than \emph{time} due to the phenomenon of line blending \cite{Andersen75aa,MeClausen07aa}, and because the RP approach does not account for differences in errorbars between individual observations. 

The results of this analysis are shown in Table\,\ref{tab:orbits}. The formal error of each parameter from the covariance matrix is included in the table to aid interpretation of the numbers. The RVs from MA97 are those obtained with a synthetic template with a line broadening of 10\kms. Table\,\ref{tab:orbits} shows that the three sets of RVs agree well; the MU01 RVs are the least good but it should be remembered that they were deliberately obtained  with a lower resolution and signal-to-noise ratio in order to mimic what \gaia\ was expected to achieve. It is also notable that the formal and MC errorbars agree very well -- formal errors are reliable in simple fits where no parameters are strongly correlated \cite{Press+92book}. The RP errors are more fragile and can either under- or over-estimate the uncertainty, so are less suitable for application to RV measurements. The published uncertainties are generally in agreement with those found here, but can sometimes underestimate the true uncertainties.

\begin{figure}[t] \centering \includegraphics[width=\textwidth]{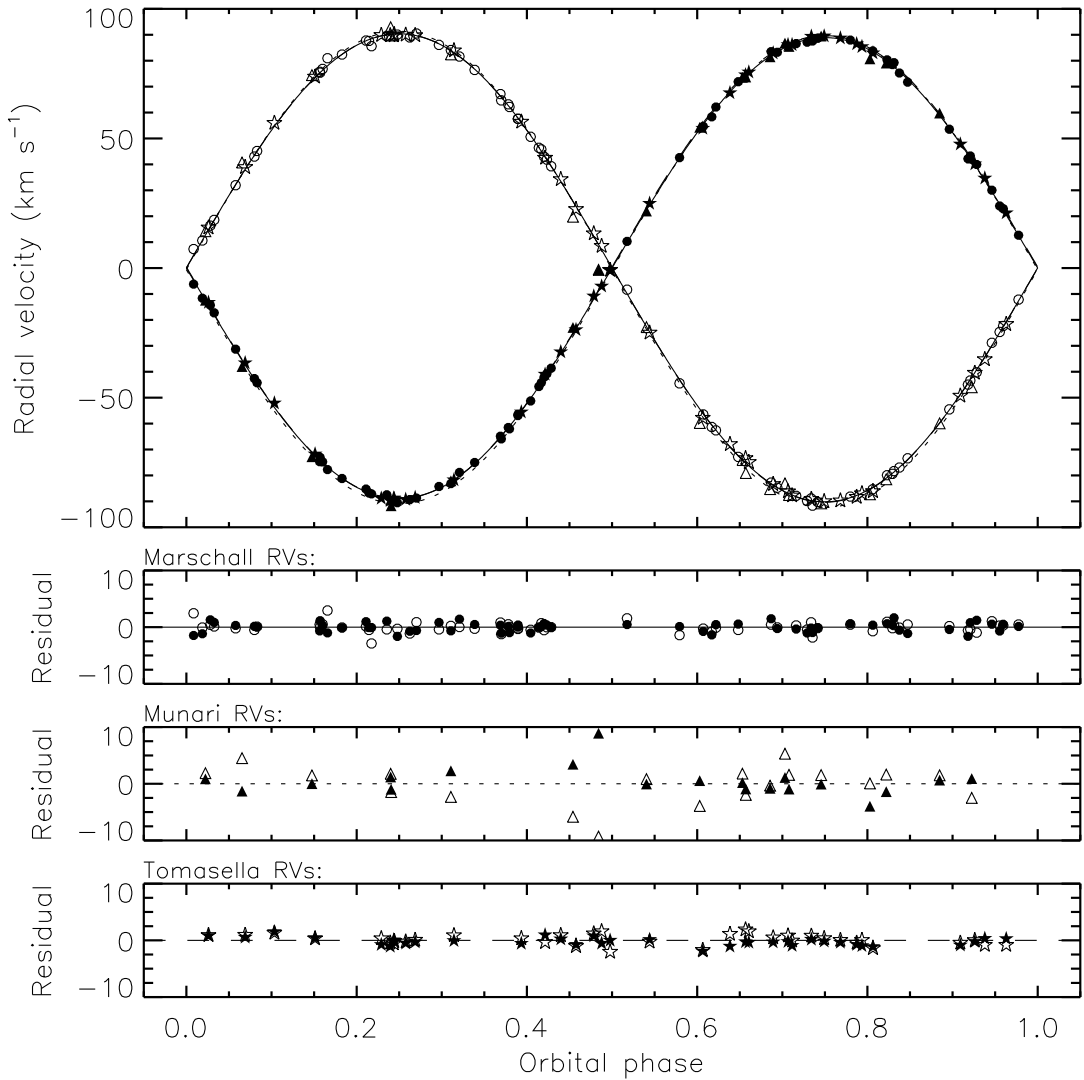} \\
\caption{\label{fig:rv} The fitted spectroscopic orbits compared to the RVs from the three different sources. 
The circles show RVs from MA97, the triangles show RVs from MU01, and the stars show RVs from TO08. In each case the RVs 
for star~A are shown using filled symbols and the RVs for star~B using open symbols. The best fits found in the 
current work for each source of RVs are shown using solid, dotted and dashed lines, respectively. The main panel
shows the RVs and fits, and the lower panels show the residuals of the fit for each source of RVs (labelled).} \end{figure}

Table\,\ref{tab:orbits} also contains the final adopted velocity amplitudes for the two stars, obtained as a weighted mean of the {\sc jktebop} results for the MA97 and TO08 RVs. Whether or not the MU01 RVs are included makes a negligible difference ($+0.03$\kms\ for $K_{\rm A}$ and $+0.01$\kms\ for $K_{\rm B}$) as they have much lower weight than the other two datasets. We did not calculate a mean value for the systemic velocities as the different datasets may not be on the same RV system and we have no use for such a value in the current work. Fig.\,\ref{fig:rv} shows the fits to all three sets of RVs. They are displayed together in the main panel but the residuals are shown separately for clarity.

% print, wmean([88.91,90.23,89.27],[0.14,0.52,0.12],chisq=chisq,sigma=sigma), chisq, sigma
% print, wmean([90.28,91.85,90.35],[0.14,1.36,0.15],chisq=chisq,sigma=sigma), chisq, sigma
%       89.1507      8.25221    0.0897437
%       90.3212      1.38711     0.102059
% print, wmean([88.91,89.27],[0.14,0.12],chisq=chisq,sigma=sigma), chisq, sigma
% print, wmean([90.28,90.35],[0.14,0.15],chisq=chisq,sigma=sigma), chisq, sigma
%       89.1175      3.81162    0.0911108
%       90.3126     0.116389     0.102348

%%%%%%%%%%%%%%%%%%%%%%%%%%%%%%%%%%%%%%%%%%%%%%%%%%%%%%%%%%%%%%%%%%%%%%%%%%%%%%%%%%%%%%%%%%%%%%%%%%%%%%%%%%%%%%%%%%%%%%%%%%%%%%%%%%%%%%%%%%%%%%%%%%%%%

\section*{Physical properties of V505 Per}

\begin{table} \centering
\caption{\em Physical properties of V505\,Per defined using the nominal solar 
units given by IAU 2015 Resolution B3 (Ref.\ \cite{Prsa+16aj}). \label{tab:absdim}}
\begin{tabular}{lr@{\,$\pm$\,}lr@{\,$\pm$\,}l}
{\em Parameter}        & \multicolumn{2}{c}{\em Star A} & \multicolumn{2}{c}{\em Star B}    \\[3pt]
Mass ratio                                  & \multicolumn{4}{c}{$0.9868 \pm 0.0016$}       \\
Semimajor axis of relative orbit (\Rsunnom) & \multicolumn{4}{c}{$14.984 \pm 0.013$}        \\
Mass (\Msunnom)                             &  1.2745 & 0.0036      &  1.2577 & 0.0030      \\
Radius (\Rsunnom)                           &  1.2941 & 0.0016      &  1.2637 & 0.0017      \\
Surface gravity ($\log$[cgs])               &  4.3194 & 0.0010      &  4.3343 & 0.0010      \\
Density ($\!$\rhosun)                       &  0.5880 & 0.0017      &  0.6232 & 0.0021      \\
Synchronous rotational velocity (\kms)      &  15.508 & 0.019       &  15.143 & 0.021       \\
Effective temperature (K)                   &    6512 & 50 $^\star$ &    6460 & 50 $^\star$ \\
Luminosity $\log(L/\Lsunnom)$               &   0.434 & 0.013       &   0.399 & 0.013       \\
$M_{\rm bol}$ (mag)                         &   3.656 & 0.033       &   3.743 & 0.034       \\
Distance (pc)                               & \multicolumn{4}{c}{$61.19 \pm 0.62$}          \\[5pt]
\end{tabular}
\newline $^\star$ Taken from TO08 but with increased errorbars.
\end{table}

The analyses above have led to final values for various parameters of the system which can be used to determine the physical properties of the stars. We did so using $r_{\rm A}$, $r_{\rm B}$, $P$ and $i$ from Table\,\ref{tab:lc}, and $K_{\rm A}$ and $K_{\rm B}$ from Table\,\ref{tab:orbits}. To these we added the \Teff\ values of the stars from TO08 (see next paragraph), the apparent magnitudes of the system given in Table\,\ref{tab:info} after converting the 2MASS magnitudes to the Johnson system, and an interstellar extinction estimate of $E(B-V) = 0.002 \pm 0.002$\,mag obtained using the {\sc stilism}\footnote{\texttt{https://stilism.obspm.fr}} online tool (Lallement \etal\ \cite{Lallement+14aa,Lallement+18aa}). These numbers were fed into the {\sc jktabsdim} code \cite{Me++05aa}, which calculates the physical properties using standard formulae \cite{Hilditch01book} and propagates uncertainties using a perturbation analysis. The results of this work are given in Table\,\ref{tab:absdim}.

For the \Teff\ measurements of the stars we adopted those from TO08, but increased the uncertainties to $\pm$50\,K as the \Teff\ scale of F-dwarfs is not currently pinned down more precisely than this \cite{Depascale+14aa,Ryabchikova+16mn,Jofre++19araa}. The surface brightness ratio determined from our modelling of the eclipses is consistent with the ratio of the \Teff s given by TO08; it also confirms to high significance that the ratio of the \Teff\ values is below unity and thus star A is hotter than star B. Detailed comparisons with theoretical predictions should account for this by comparing the \Teff\ of star A and the ratio of the \Teff s rather than the two \Teff\ values directly \cite{MeClausen07aa}.

% IDL> print, [0.0036,0.0030,0.0016,0.0017]*100.0d0/[1.2745,1.2577,1.2941,1.2637] 
%       0.28246370      0.23853066      0.12363804      0.13452560

The properties of the system are measured to an exceptionally high precision: 0.26\% in mass and 0.13\% in radius. The excellence of these results can be attributed to the availability of multiple sets of high-quality RVs and the remarkable light curve obtained using \tess. Only a few other EBs have properties measured to a comparable precision (e.g.\ AI\,Phe \cite{Maxted+20mn} and FM\,Leo \cite{Graczyk+21aa}). We determined the distance to the system using the calibration of $K$-band surface brightness versus \Teff\ presented by Kervella \etal\ \cite{Kervella+04aa}, finding $61.19 \pm 0.62$\,pc. This is slightly below the distance of $62.23 \pm 0.08$\,pc found from the parallax of the system in \gaia\ EDR3. We note that Ref. \cite{Bailerjones+21aj} obtained a distance of $62.03 \pm 0.10$\,pc from their re-interpretation of the \gaia\ EDR3 parallaxes using priors from a three-dimensional model of the Milky Way. The dominant contributor to the uncertainty in our distance measurement is the $K_s$-band apparent magnitude from 2MASS.

Although the two stars are very similar, their masses and radii differ by much more than the measurement errors so a comparison with the predictions of theoretical stellar evolutionary models is of interest. For this we chose the PARSEC models \cite{Bressan+12mn} and overlaid their predicted properties on the observed ones in the mass--radius and mass--\Teff\ diagrams. We found a good fit to all properties for a fractional metal abundance of $Z = 0.017$ and an age of $1050 \pm 50$\,Myr. Predictions for $Z=0.014$ or $Z=0.020$ significantly over- or under-predict the measured \Teff\ values so can be ruled out. 

TO08 measured the metallicities of both stars to be [M/H] $= -0.12 \pm 0.03$ via $\chi^2$ fitting synthetic spectra to the observed spectra. The heavy-element mixture adopted for the PARSEC models equates to a solar value of $Z_\odot = 0.01524$ so the measured [M/H] corresponds to $Z=0.0116$. This conflicts with the results of the comparison in the mass--radius and mass--\Teff\ diagrams, suggesting that a reappraisal of the photospheric chemical composition is warranted.

%%%%%%%%%%%%%%%%%%%%%%%%%%%%%%%%%%%%%%%%%%%%%%%%%%%%%%%%%%%%%%%%%%%%%%%%%%%%%%%%%%%%%%%%%%%%%%%%%%%%%%%%%%%%%%%%%%%%%%%%%%%%%%%%%%%%%%%%%%%%%%%%%%%%%

\section*{Summary}

V505\,Per is a dEB containing two F5\,V stars on a 4.22\,d circular orbit. Time-series photometry and RVs have been presented and analysed in three publications, and a light curve from the \tess\ satellite has recently become available. We determined the physical properties of the system based on the \tess\ data and the published RVs. We measured the masses to a precision of 0.26\% and the radii to a precision of 0.13\%. Analysis of the existing data for the system led to the conclusion that formal errors can be trusted for RVs, where correlations between parameters are weak, but not for light curves, where parameter correlations are often strong. Including the precise \Teff\ and metallicity values from TO08, V505\,Per becomes one of the dEBs with the most precisely determined basic physical properties.

We find that the PARSEC theoretical stellar evolutionary models provide a good match to the measured masses, radii and \Teff\ values for an age of approximately 1\,Gyr and a modestly supersolar metal abundance. As TO08 found both stars to have a slightly subsolar metallicity, we conclude that a detailed spectroscopic chemical abundance analysis should be performed for this system.

%%%%%%%%%%%%%%%%%%%%%%%%%%%%%%%%%%%%%%%%%%%%%%%%%%%%%%%%%%%%%%%%%%%%%%%%%%%%%%%%%%%%%%%%%%%%%%%%%%%%%%%%%%%%%%%%%%%%%%%%%%%%%%%%%%%%%%%%%%%%%%%%%%%%%

\section*{Acknowledgements}

We thank Lina Tomasella, Claud Lacy and Guillermo Torres for their help in trying to find the elusive light curves published by MA97.
We also acknowledge a helpful report from an ananymous referee and useful comments from Dariusz Graczyk.
This paper includes data collected by the \tess\ mission. Funding for the \tess\ mission is provided by the NASA's Science Mission Directorate.
The following resources were used in the course of this work: the NASA Astrophysics Data System; the SIMBAD database operated at CDS, Strasbourg, France; and the ar$\chi$iv scientific paper preprint service operated by Cornell University.
I acknowledge that this work may give the impression that I am obsessed with errorbars.

%%%%%%%%%%%%%%%%%%%%%%%%%%%%%%%%%%%%%%%%%%%%%%%%%%%%%%%%%%%%%%%%%%%%%%%%%%%%%%%%%%%%%%%%%%%%%%%%%%%%%%%%%%%%%%%%%%%%%%%%%%%%%%%%%%%%%%%%%%%%%%%%%%%%%

% \bibliographystyle{obsmaga}
% \bibliography{jkt}

%%%%%%%%%%%%%%%%%%%%%%%%%%%%%%%%%%%%%%%%%%%%%%%%%%%%%%%%%%%%%%%%%%%%%%%%%%%%%%%%%%%%%%%%%%%%%%%%%%%%%%%%%%%%%%%%%%%%%%%%%%%%%%%%%%%%%%%%%%%%%%%%%%%%%
\end{document}